\documentclass[prb,twocolumn,showpacs]{revtex4}

\usepackage{graphicx}

\begin{document}  

\title{Orbital Lamb shift and mixing of 
the pseudo-zero-mode Landau levels\\
in $ABC$-stacked trilayer graphene
}
\author{K. Shizuya}
\affiliation{Yukawa Institute for Theoretical Physics\\
Kyoto University,~Kyoto 606-8502,~Japan }

\begin{abstract}

In a magnetic field graphene trilayers  support a characteristic multiplet 
of 12 zero(-energy)-mode Landau levels 
with a threefold degeneracy in Landau orbitals. 
It was earlier noted for bilayer graphene that Coulombic vacuum fluctuations,
specific to graphene, lift the orbital degeneracy of such zero-energy modes
and that these $\lq\lq$Lamb-shfted" orbital modes, with filling, get mixed 
via the Coulomb interaction.
It is pointed out that analogous orbital Lamb shift and mixing 
of zero-mode levels can also take place,
with an enriched symmetry content, in $ABC$-stacked trilayer graphene;
and its consequences are discussed in the light of experimental results. 

\end{abstract} 

\pacs{73.22.Pr,73.43.-f,75.25.Dk}

\maketitle

\section{Introduction}

Graphene,~\cite{NG,ZTSK,ZA}
an atomic layer of graphite that supports massless Dirac fermions, 
attracts great attention 
for its unique and promising electronic properties.
Recently interest appears to center on 
bilayers and few layers of graphene,
where the added layer degree of freedom makes 
the physics and applications of graphene richer. 
In particular, bilayer graphene and some types of multilayers
enjoy the property that their band gaps 
are externally tunable.~\cite{MF,OBSHR,Mc,CNMPL} 

A notable signal of Dirac fermions is
the fact that  graphene, in a magnetic field, supports
a characteristic set of zero-energy Landau levels, 
whose emergence and degeneracy 
have a topological origin in the chiral anomaly.~\cite{NS}
Monolayer graphene has four such zero-energy levels owing to 
the spin and valley degeneracy, 
and they are responsible 
for the observed half-integer quantum Hall effect.\cite{NG,ZTSK}
In bilayer graphene there are eight such levels, 
with an extra twofold degeneracy~\cite{MF} 
in Landau orbitals $n$=0 and 1.  
This $\lq \lq$orbital" degeneracy is a consequence of 
topology and the added layer, 
and $N$-layer graphene necessarily has $4N$ zero-energy Landau levels
with $N$-fold orbital degeneracy.
In the presence of Zeeman coupling, Coulomb interactions, etc.,
these zero-energy levels 
evolve into a variety of pseudo-zero-mode (PZM) levels, 
or broken-symmetry states, as discussed theoretically.~\cite{BCNM, NL,GGJM}
The interplay of orbital degeneracy and Coulomb interactions brings 
about a new realm of 
quantum phenomena~\cite{BCNM,KSpzm,BCLM,CLBM,CLPBM,CFL} 
in the PZM sector, 
such as orbital mixing and orbital-pseudospin waves.

Graphene is distinguished from conventional electron systems 
by the feature that 
it is an intrinsically many-body system equipped with 
the quantum vacuum, or the valence band acting as the Dirac sea.
Quantum fluctuations of the Dirac sea are sizable,
even leading to ultraviolet divergences;
and one encounters such field-theoretic (or many-body) phenomena 
as velocity renormalization,~\cite{velrenorm}
screening of charge,~\cite{KSbgr} 
and nontrivial Coulombic corrections to 
cyclotron resonance.~\cite{JHT,IWFB,BMgr,KCC,KScr}
Quantum fluctuations also affect the PZM levels substantially. 
They work to lift~\cite{KSLs} the orbital degeneracy of the PZM levels 
in bilayer graphene;
each orbital mode responds to quantum fluctuations differently 
and gets shifted, just like the Lamb shift~\cite{Lambshift} in the hydrogen atom, 
where the field-theoretic effect of quantum electrodynamics
was revealed for the first time historically. 
The Lamb-shifted orbital modes get mixed 
via Coulomb interactions
and govern the fine structure of the PZM sector.

A number of recent  
experiments~\cite{BZZL,KEPF,TWTJ,LLMC,BJVL,ZZCK,JCST}
have verified that 
the electronic properties of graphene trilayers
strongly depend on the stacking order, 
with Bernal $(ABA)$-stacked trilayers remaining metallic 
in contrast to rhombohedral $(ABC$-stacked) trilayers 
which exhibit a tunable band gap.
Actually trilayers drew theorists' attention~\cite{GCP,KA,NCGP,KM80} 
even before experiments
and their rich electronic properties~\cite{KMc81,ZSMM,YRK} 
have been under active study.
Currently considerable attention~\cite{ZTM, BCR,CRGB} is directed to 
$ABC$-stacked trilayers which are
a chiral generalization of bilayer graphene.
In view of this, it is of interest to ask 
how the Coulombic vacuum and orbital dynamics
generalizes to trilayers.

The purpose of this paper is to examine the effect of 
Coulombic vacuum fluctuations 
in trilayers and show that the orbital Lamb shift 
and orbital mixing of the PZM levels are also present,
with an enriched symmetry content, in $ABC$-stacked trilayer graphene.
It is noted, in particular, that level mixing takes place without level crossing;
this mechanism would, for high-quality samples, lead to an observable
sequence of fully-split broken-symmetry quantum Hall states
in the PZM sector.

In Sec.~II we briefly review some basic features of 
the PZM levels in $ABC$-trilayer graphene, 
and in Sec.~III show that vacuum fluctuations 
lift their orbital degeneracy.
In Sec.~IV  we discuss in a simplified setting 
how orbital mixing of the PZM levels takes place 
via the Coulomb interaction.
In Sec.~V we examine the hierarchy of broken-symmetry states 
under practical conditions. 
Section~VI is devoted to a summary and discussion.

\section{trilayer graphene}

The $ABC$-stacked trilayer graphene consists of three graphene layers 
with vertically-arranged dimer bonds  $(B_{1}, A_{2})$ and $(B_{2}, A_{3})$,
where $(A_{i}, B_{i})$ 
denote inequivalent lattice sites in the $i$-th layer.
The interlayer coupling $\gamma_{0}\equiv \gamma_{B_{i}A_{i}} \sim 3$\, eV 
is related to the Fermi velocity 
$v = (\sqrt{3}/2) a_{L}\gamma_{0}/\hbar \sim 10^6$ m/s in monolayer graphene.
Interlayer hopping via the nearest-neighbor dimer coupling~\cite{Malard} 
$\gamma_{1}\equiv \gamma_{B_{1}A_{2}} 
=\gamma_{B_{2}A_{3}} \sim $ 0.4\, eV
leads to soft cubic spectra~\cite{GCP} $\propto |{\bf p}|^3$ 
in the low-energy branches $|\epsilon|< \gamma_{1}$.

The effective Hamiltonian for $ABC$-stacked trilayer graphene 
with such intralayer and interlayer couplings 
is written as~\cite{KM80}
\begin{eqnarray}
H^{\rm tri} &=&\!\! \int\! d^{2}{\bf x}\, 
\Big[ (\Psi^{K})^{\dag}\, {\cal H}_{K} \Psi^{K}
+ (\Psi^{K'})^{\dag}\, {\cal H}_{K'}\, \Psi^{K'}\Big], \nonumber\\
{\cal H}_{K} &=& \left(
\begin{array}{lll}
D_{1}&  V &  W \\
 V^{\dag} & D_{2}& V\\
W^{\dag}& V^{\dag} & D_{3} \\
\end{array}
\right),
\nonumber\\
D_{i} &=& \left(
\begin{array}{cc}
U_{i}&  v\, p^{\dag} \\
 v\, p & U_{i} \\
\end{array}
\right),\ 
V = \left(
\begin{array}{cc}
-v_{4}\, p^{\dag}&  v_{3}\, p \\
 \gamma_{1} & -v_{4}\, p^{\dag} \\
\end{array}
\right),
\nonumber\\
W&=& \left(
\begin{array}{cc}
0& \gamma_{2}/2 \\
0 & 0\\
\end{array}
\right),
\label{Htri}
\end{eqnarray}
with 
$p= p_{x} + i p_{y}$ and $p^{\dag}= p_{x} - i p_{y}$.
Here  $\Psi^{K}=(\psi_{A_{1}}, \psi_{B_{1}}, \psi_{A_{2}},\psi_{B_{2}},
\psi_{A_{3}}, \psi_{B_{3}})^{\rm t}$
stands for the electron field at the $K$ valley.
$v_{3}$ and $v_{4}$ are related to the nonleading interlayer couplings
$\gamma_{3} \equiv \gamma_{A_{1} B_{2}}$ and
$\gamma_{4} \equiv \gamma_{A_{1} A_{2}}$, respectively, 
and $\gamma_{2} \equiv \gamma_{A_{1} B_{3}}$.
$(U_{1}, U_{2}, U_{3})$ stand for the on-site energies of the three layers;
we take $U_{2}=0$ without loss of generality. 
As in bilayer graphene,~\cite{MF} 
these biases $\{U_{i} \}$
open a tunable band gap~\cite{GCP} $\sim U_{1}-U_{3}$.
${\cal H}_{K}$ is diagonal in (suppressed) electron spin.

The Hamiltonian ${\cal H}_{K'}$ at another valley is given by ${\cal H}_{K}$ 
with $p \rightarrow  - p_{x} + i p_{y} = -p^{\dag}$
and $p^{\dag}\rightarrow - p$,
and acts on a spinor of the same sublattice content as $\Psi^{K}$.
Actually, ${\cal H}_{K'}$ is unitarily equivalent to 
${\cal H}_{K}$ with the sign of $v_{3}$ and $\gamma_{2}$ reversed
and with layer 1 and layer 3 interchanged, 
\begin{eqnarray}
S^{\dag}{\cal H}_{K'}S  &=& {\cal H}_{K}|_{-v_{3}, -\gamma_{2}; 
U_{1} \leftrightarrow U_{3} } 
\label{Hequi},  \nonumber\\
S &=& \left(
\begin{array}{ccc}
&  & \sigma_{2}\\
& -\sigma_{2} & \\
\sigma_{2} & & \\
\end{array}
\right).
\label{SHS}
\end{eqnarray}
In view of this, we adopt 
$\hat{\cal H}_{K'} =S^{\dag} {\cal H}_{K'} S$ for ${\cal H}_{K'}$
and simply pass to the $K'$ valley by reversing the sign of 
$v_{3}$ and $\gamma_{2}$ 
and interchanging $U_{1}$ and $U_{3}$
in the $K$-valley expressions.
Nonzero $v_{3}$, $\gamma_{2}$  and bias $U_{1} - U_{3}$
thus act as valley-symmetry breakings.
Remember that in this representation $\hat{\cal H}_{K'} $ acts 
on a spinor of the form $\Psi^{K'}=(\psi_{B_{3}}, \psi_{A_{3}}, \psi_{B_{2}},\psi_{A_{2}},
\psi_{B_{1}}, \psi_{A_{1}})^{\rm t}$.

A direct link between the $K$- and $K'$-valley representations, 
such as Eq.~(\ref{SHS}),
was also noted~\cite{KSLs} for bilayer graphene.
We remark that such a link is not shared by $ABA$-stacked trilayers, 
where the Landau-level spectra significantly differ~\cite{KMc81} 
between the two valleys for nonzero biases 
(though they coincide for zero bias). 

We have discussed the general structure of trilayer parameters 
for completeness.
For our present analysis of quantum effects in $ABC$ trilayers 
we retain only the leading parameters $(v, \gamma_{1}, U_{i})$;
the effect of nonleading couplings $(v_{3}, v_{4}, \gamma_{2})$ 
is discussed later in Sec.~V. 
In addition,  we focus on the case of a symmetric bias~\cite{GCP} 
by choosing $U_{3}= -U_{1} \equiv u/2$.

Let us place trilayer graphene in a strong uniform magnetic field 
$B_{z} = B>0$ normal to the sample plane;
we set, in ${\cal H}_{K}$, 
$p\rightarrow \Pi = p + eA$
with $A= A_{x}+ iA_{y}= -B\, y$, 
and denote the the magnetic length as $\ell=1/\sqrt{eB}$;
setting $a \equiv \sqrt{2eB}\,  \Pi^{\dag}$ then yields $[a,a^{\dag}]=1$.
It is easily seen that the eigenmodes of ${\cal H}_{K}$ have the structure 
\begin{eqnarray}
\Psi_{n} &=& \Big( |n -3 \rangle\, b_{n}^{(1)} ,|n -2 \rangle\, d_{n}^{(1)} ,
|n -2 \rangle\, b_{n}^{(2)} , \nonumber\\
&&
|n -1 \rangle\,  d_{n}^{(2)} ,  |n -1 \rangle\, b_{n}^{(3)} , 
|n \rangle\, d_{n}^{(3)}  \Big)^{\rm t}
\end{eqnarray} 
with $n=0,1,2,\cdots$, where only the orbital eigenmodes are shown 
using the standard harmonic-oscillator basis $\{ |n\rangle \}$
(with the understanding that $|n\rangle =0$ for $n<0$).
The coefficients 
${\bf v}_{n}=(b_{n}^{(1)}, d_{n}^{(1)}, b_{n}^{(2)}, d_{n}^{(2)}, 
b_{n}^{(3)}, d_{n}^{(3)} )^{\rm t}$
 for each $n=3,4, \dots$ are given by the eigenvectors 
 (chosen to form an orthonormal basis)  of the reduced Hamiltonian
$\hat{\cal H}_{\rm red} \equiv \omega_{c} {\cal H}_{n}$ 
with
\begin{equation}
{\cal H}_{n} = 
\left(
\begin{array}{cccccc}
-M &  \sqrt{n-2}   & & & & \\
 \sqrt{n-2}  &-M &\hat{\gamma} & & &\\
& \hat{\gamma} & 0 & \sqrt{n - 1}  & & \\
& & \sqrt{n -1} & 0 & \hat{\gamma} &   \\
& & & \hat{\gamma} & M & \sqrt{n}     \\
& &  &   & \sqrt{n}  & M \\
\end{array}
\right),
\nonumber\\
\label{Hn}
\end{equation}
where 
\begin{equation}
\omega_{c}\equiv \sqrt{2}\, v/\ell 
\approx 36.3 \times v[10^{6}{\rm m/s}]\, \sqrt{B[{\rm T}]}\ {\rm meV},
\end{equation} 
with $v$ in units of $10^{6}$m/s and $B$ in tesla, 
is the characteristic cyclotron energy 
for monolayer graphene;
$M \equiv {1\over{2}}\, u/\omega_{c}$ and  
$\hat{\gamma} \equiv \gamma_{1}/\omega_{c}$.
Note that eigenvectors ${\bf v}_{n}$ can be taken real since 
${\cal H}_{n}$ is a real symmetric matrix.

Solving the secular equation shows that there are 6 branches of 
Landau levels for each integer $n\ge 3$.  
We denote the eigenvalues as
$\epsilon_{-n''} < \epsilon_{-n'}<\epsilon_{-n}<0<\epsilon_{n}<\epsilon_{n'}
< \epsilon_{n''}$,
so that the index $\pm n$ reflects the sign of $\epsilon_{n}$;
$|\epsilon_{\pm n'}| \gtrsim \gamma_{1}$ 
and $|\epsilon_{\pm n''}|  \gtrsim \gamma_{1}$. 
The $|n|=3$ levels, e.g.,  consist of the $n=(\pm 3. \pm 3',\pm 3'')$ branches.

There are also solutions for $n=$2, 1 and 0, for which 
${\cal H}_{n}$ is reduced to a matrix of smaller rank 5, 3 and 1.
For $n=0$, $\hat{\cal H}_{\rm red}$ has an obvious eigenvalue 
$\epsilon_{0}=U_{3}= u/2$ with eigenvector
${\bf v}_{0} = (0,0,0,0,0,1)^{\rm t}$
or 
\begin{equation}
\Psi_{0} = ( 0,0,0,0,0, |0\rangle)^{\rm t}.
\label{psizero}
\end{equation}
For $n=1$, $\hat{\cal H}_{\rm red}$ has three eigenvalues
$(\epsilon_{1}, \epsilon_{\pm 1'})$, 
which, for $u=0$,
read $(0, \pm \sqrt{ \hat{\gamma}^2 +1})\, \omega_{c}$.
The zero-energy solution, in particular, takes the form:
\begin{eqnarray}
\Psi_{1} \stackrel{u=0}{=}  
c_{1}\, \big(0,0,0,- \kappa\, |0\rangle, 0,  |1\rangle \big)^{\rm t},
\label{psione}
\end{eqnarray}
with $\kappa \equiv 1/\hat{\gamma}$ 
and $c_{1}= \hat{\gamma}/\sqrt{\hat{\gamma}^2 +1} = 1/\sqrt{1+ \kappa^2}$.
For $n=2$, $\hat{\cal H}_{\rm red}$ has five eigenvalues 
$(\epsilon_{2}, \epsilon_{\pm 2'}, \epsilon_{\pm 2''})$,
with  $\epsilon_{2} =0$ and 
$ |\epsilon_{\pm 2'}| \sim |\epsilon_{\pm 2''}| \sim \gamma_{1}$
for $u=0$.
The zero-energy solution takes the form 
\begin{equation}
\Psi_{2} \stackrel{u=0}{=} 
c_{2}\, \big( 0, \sqrt{2}\,\kappa^2\, |0\rangle, 
0, -\sqrt{2}\,\kappa\,  |1\rangle, 0, |2\rangle \big)^{\rm t},
\label{psitwo}
\end{equation}
with $c_{2}= \hat{\gamma}^2/\sqrt{2+ 2\, \hat{\gamma}^2 + \hat{\gamma}^4} 
=1/ \sqrt{1+  2 \kappa^2 + 2 \kappa^4}$.
Note that these zero-energy solutions $(\Psi_{0},\Psi_{1},\Psi_{2})$
reside predominantly on the $B_{3}$ lattice sites of the third layer;
correspondingly, the zero-energy solutions at the $K'$ valley 
reside predominantly on the $A_{1}$ sites of the first layer.

Of our particular concern are these three zero-energy modes
$(\Psi_{0},\Psi_{1},\Psi_{2})$.
For $u=0$ there are 12 such zero-energy Landau levels 
differing in spin, valley and orbital $[n=(0,1,2)]$ degrees of freedom;
their presence is dictated 
by the nonzero index~\cite{NS, KSbgr} 
of the Dirac Hamiltonian ${\cal H}_{K} \oplus {\cal H}_{K'}$ 
with only  $v$ and $\gamma_{1}$ retained.

For nonzero bias $u\not=0$ they evolve into the pseudo-zero modes 
with nonzero energies,
\begin{eqnarray}
&& (\epsilon^{u}_{0}, \epsilon^{u}_{1} , \epsilon^{u}_{2} ) 
= ( 1, 1-z_{1}, 1- z_{2})\, u/2,  \nonumber\\
&& z_{1} = \kappa^2 (c_{1})^2 + O(\hat{u}^2 \kappa^6), \nonumber\\
&& z_{2} = 2 \kappa^2 (1+ 2\kappa^2) (c_{2})^2 + O(\hat{u}^2\kappa^4), 
\end{eqnarray}
where $\hat{u}\equiv u/\omega_{c}$.
One can also write 
$\epsilon^{u}_{1} \approx  (c_{1})^2\,  u/2$ and 
$\epsilon^{u}_{2}
\approx  (1- 2\kappa^4) (c_{2})^2\, u/2$.

For a numerical estimate let us take, as typical values,~\cite{Malard} 
$\gamma_{0}=3.16$ eV  (or $v\approx 1.0 \times 10^6$ m/s) 
and $\gamma_{1}=0.4$ eV.
They yield 
$\hat{\gamma} \approx 3.41$ and $\kappa \approx 0.293$ at $B=$ 10T, 
which in turn lead to
$c_{1} \approx 0.960$, $c_{2} \approx 0.918$, 
$z_{1} \approx 0.08$ and $z_{2} \approx 0.17$. 
One thus has 
\begin{eqnarray}
(\epsilon_{0}^{u}, \epsilon_{1}^{u}, \epsilon_{2}^{u}) 
\approx (1, 0.92, 0.83)\, u/2
\end{eqnarray}
for $u \ll \omega_{c}$.

One can pass to the $K'$ valley by  setting $u \rightarrow - u$ 
in the $K$-valley expressions.
The eigensystems $(\epsilon_{n}, {\bf v}_{n})$ at the two valleys 
are related as
\begin{eqnarray}
&&\epsilon_{n}|_{K'} = -\epsilon_{-n}|_{K}, \nonumber\\  
&&b_{n}^{(i)}|_{K'} = - b_{-n}^{(i)}|_{K}, d_{n}^{(i)}|_{K'} = d_{-n}^{(i)}|_{K}
\label{eigensystems}
\end{eqnarray}
for each mode $(n, n', n'')$ and $i \in (1,2,3)$.
For later convenience, we continue to use $n=(0,1, 2)$ 
to specify the PZM levels at the $K'$ valley;
one can thus effectively set $n=\pm 0\rightarrow 0$,  
$\pm 1\rightarrow 1$ and $\pm 2\rightarrow 2$.
When the interlayer bias $u$ is turned on, 
these PZM levels go up or down oppositely at the two valleys,
opening a band gap $\sim u$.

The Landau-level structure is made explicit by passing to
the $|n,y_{0}\rangle$ basis  (with $y_{0}\equiv \ell^{2}p_{x}$) 
via the expansion
$(\Psi^{K} ({\bf x}), \Psi^{K'} ({\bf x}) ) 
= \sum_{n, y_{0}} \langle {\bf x}| n, y_{0}\rangle\, \{ \psi^{n;a}_{\alpha}(y_{0})\}$, 
where $n$ refers to the level index, $\alpha \in (\uparrow, \downarrow)$ 
to the spin, and $a \in (K,K')$ to the valley.
The charge density 
$\rho_{-{\bf p}} =\int d^{2}{\bf x}\,  e^{i {\bf p\cdot x}}\,\rho$ 
with $\rho = (\Psi^{K})^{\dag}\Psi^{K} +  (\Psi^{K'})^{\dag}\Psi^{K'}$ 
is thereby written as~\cite{KSLs}
\begin{eqnarray}
\rho_{-{\bf p}} &=& \gamma_{\bf p}\sum_{k, n =-\infty}^{\infty}
\sum_{a,\alpha} g^{k n;a}_{\bf p}\, 
R^{k n;aa}_{\alpha\alpha;\bf p}, \nonumber\\
R^{kn;ab}_{\alpha\beta;{\bf p}}&\equiv& \int dy_{0}\,
{\psi^{k,a}_{\alpha}}^{\dag}(y_{0})\, e^{i{\bf p\cdot r}}\,
\psi^{n,b}_{\beta} (y_{0}),
\label{chargeoperator}
\end{eqnarray}
where $\gamma_{\bf p} \equiv  e^{- \ell^{2} {\bf p}^{2}/4}$; 
${\bf r} = (i\ell^{2}\partial/\partial y_{0}, y_{0})$
stands for the center coordinate with uncertainty 
$[r_{x}, r_{y}] =i\ell^{2}$;
the level sum $\sum_{n}$ is taken over possible $(n,n',n'')$.

The coefficient matrix $g^{k n;a}_{\bf p} \equiv g^{kn}_{\bf p}|_{a}$ 
at valley $a\in (K,K')$ is constructed from  the eigenvectors ${\bf v}_{n}|_{a}$,
\begin{eqnarray}
g^{kn}_{\bf p} &=& b_{k}^{(1)} b_{n}^{(1)}\, f_{\bf p}^{|k|-3,|n|-3} \nonumber\\
&& +( d_{k}^{(1)} d_{n}^{(1)} + b_{k}^{(2)} b_{n}^{(2)})\, f_{\bf p}^{|k|-2,|n|-2} 
\nonumber\\
&& + (d_{k}^{(2)} d_{n}^{(2)}+ b_{k}^{(3)} b_{n}^{(3)})\, f_{\bf p}^{|k|-1,|n|-1} 
\nonumber\\
&& + d_{k}^{(3)} d_{n}^{(3)}\, f_{\bf p}^{|k|,|n|},
\label{gkn}
\end{eqnarray}
where
\begin{equation}
f^{k n}_{\bf p} 
= \sqrt{n!/k!}\, (-\bar{q}/\sqrt{2})^{k-n}\, L^{(k-n)}_{n} (|\bar{q}|^{2}/2)
\label{fknp}
\end{equation}
for $k \ge n\ge0$, and $f^{n k}_{\bf p} = (f^{k n}_{\bf -p})^{\dag}$;
$\bar{q} =\ell (p_{x}\! -i\, p_{y})$; it is understood that 
$f^{kn}_{\bf p}=0$ for $k<0$ or $n<0$.
As seen from Eq.~(\ref{eigensystems}), $g^{k n;a}_{\bf p}$ 
at the two valleys are related as 
\begin{equation}
g^{mn}_{\bf p}|_{K'}=g^{-m,-n}_{\bf p}|_{K},\ \ 
g^{mn;a}_{\bf p}|_{u}=g^{-m,-n;a}_{\bf p}|_{-u}.
\label{gmnproperty}
\end{equation}

Within the $n\in (0,1,2)$ sector, 
$g^{k n;a}_{\bf p}$ are functions~\cite{fngz}
of $(\hat{u}^2, \kappa^2)$ 
and are thus common to both valleys; 
for $u=0$, they read
\begin{eqnarray}
g^{00}_{\bf p} &=& 1,\ \  g^{01}_{\bf p} = c_{1} \ell\, p/\sqrt{2},\ \ 
g^{10}_{\bf p} = -c_{1} \ell\, \bar{p}/\sqrt{2},\nonumber\\
g^{02}_{\bf p} &=& c_{2}\,  \ell^{2} p^2/(2\sqrt{2}),\nonumber\\
g^{11}_{\bf p} &=&1- (c_{1})^{2}\, \textstyle{1\over{2}}\ell^{2}{\bf p}^{2}, \ \ 
g^{12}_{\bf p} =  \lambda_{\bf p}\, g^{01}_{\bf p}, \nonumber\\
g^{22}_{\bf p} &=& 1- (c_{2})^2
\big[ (1+\kappa^2)\, \ell^{2}{\bf p}^2 
- \textstyle{1\over{8}}(\ell^{2}{\bf p}^2)^2\big], \nonumber\\
\lambda_{\bf p} &=& \sqrt{2} \,
( 1 + \kappa^2 - \textstyle{1\over{4}}  \ell^2\, {\bf p}^2)\, c_{2},
\end{eqnarray}
with $c_{1}= 1/\sqrt{1+ \kappa^2}$ and
$c_{2} =1/ \sqrt{1+  2 \kappa^2 + 2 \kappa^4}$.

From now on we frequently suppress
summations over levels $n$, spins $\alpha$ and valleys $a$, 
with the convention that the sum is taken over repeated indices.
The Hamiltonian $H^{\rm bi}$ projected to 
the PZM levels is thereby written as
\begin{equation}
H_{u} = \epsilon^{u}_{n}\, \delta R^{nn}_{\beta\beta;{\bf 0}} 
- \mu_{\rm Z}\, (T_{3})_{\beta\alpha} R^{nn;aa}_{\alpha\beta;{\bf 0}} 
\label{Hzero}
\end{equation}
with $n\in (0,1,2)$ and
$\delta R^{mn}_{\alpha\beta;{\bf 0}} \equiv R^{mn;KK}_{\alpha\beta;{\bf 0}}  
-R^{mn;K'K'}_{\alpha\beta;{\bf 0}}$. 
Here the Zeeman term $\mu_{\rm Z} \equiv g^{*}\mu_{\rm B}B
\approx 0.12\, B[{\rm T}]$ meV is introduced 
via the spin matrix 
$T_{3} = \sigma_{3}/2$.

\section{vacuum fluctuations}

In this section we examine the effect of Coulombic quantum fluctuations 
on the PZM multiplet.
The Coulomb interaction is written as
\begin{equation}
V
= {1\over{2}} \sum_{\bf p}
v_{\bf p}\, :\rho_{\bf -p}\, \rho_{\bf p}:,
\label{Hcoul}
\end{equation}
where
$v_{\bf p}= 2\pi \alpha/(\epsilon_{\rm b} |{\bf p}|)$ 
with 
$\alpha = e^{2}/(4 \pi \epsilon_{0}) \approx 1/137$ and 
the substrate dielectric constant $\epsilon_{\rm b}$;
$\sum_{\bf p} =\int d^{2}{\bf p}/(2\pi)^{2}$.   
For simplicity we ignore the difference 
between the intralayer and interlayer Coulomb potentials.

In this paper we generally focus on many-body ground states $|G\rangle$ 
with a homogeneous density, realized at integer filling factor $\nu \in [-6, 6]$.
We set the expectation values  
$\langle G| R^{mn;ab}_{\alpha\beta; {\bf k}}|G \rangle 
= \delta_{\bf k,0}\, \rho_{0}\,  \nu^{mn;ab}_{\alpha\beta}$
with $\rho_{0} = 1/(2\pi \ell^{2})$ and 
$\delta_{\bf k,0}= (2\pi)^2\, \delta^2({\bf k})$;
accordingly, the filling factor $\nu^{nn;aa}_{\alpha\alpha}=1$ 
for a filled level specified by $(n,a,\alpha)$.

Let us define the Dirac sea $|{\rm DS}\rangle$ as the valence band 
with levels below the PZM sector 
(i.e., levels with $n\le -3$, $n'\le -1'$ and $n''\le -2''$) all filled.
We construct the Hartree-Fock Hamiltonian
$V^{\rm HF}$ out of $V$ as  the effective Hamiltonian 
that governs the electron states over $|{\rm DS}\rangle$.
Let us write $V^{\rm HF} = V_{\rm D}+ V_{\rm X}$. 
As usual, the direct interaction
$V_{\rm D}
\propto v_{\bf p\rightarrow 0}\, R^{m'm';bb}_{\beta\beta; {\bf 0}}$
is removed if one takes into account neutralizing
positive background charges.  
We thus focus on the exchange interaction
\begin{equation}
V_{\rm X}
= - \sum_{\bf p}v_{\bf p}\gamma_{\bf p}^{2}\,
g^{m n';b}_{\bf -p}\,g^{m'n;a}_{\bf p}\,
\nu^{mn;ba}_{\beta \alpha}\, R^{m' n';ab}_{\alpha\beta;{\bf 0}},
\label{Vex}
\end{equation}
where we sum over filled levels $(m,n)$ 
and retain the PZM sector $m',n'\in (0,1,2)$.

Let us first extract, out of $V_{\rm X}$,
the contribution from the Dirac sea,
\begin{equation}
V^{\rm DS}_{\rm X}
= - \sum_{\bf p}v_{\bf p}\gamma_{\bf p}^{2}\,
\sum_{n\in {\rm DS}}|g^{m'n;a}_{\bf p}|^2\, 
R^{m' m';aa}_{\alpha\alpha;{\bf 0}},
\label{VxDS}
\end{equation}
where the sum over $m'\in (0,1,2)$, $a\in (K,K')$ 
and $\alpha \in (\uparrow, \downarrow)$ is understood.
Actually, the sum over infinitely many filled levels 
with $-\infty < n \in {\rm DS}$
gives rise to  an ultraviolet divergence.

Fortunately one can isolate the divergence 
and even evaluate $V^{\rm DS}_{\rm X}$ exactly 
for zero bias $u \rightarrow 0$, 
as done for the bilayer case.~\cite{KSLs}
Note first that, as seen from Eq.~(\ref{gmnproperty}),
$g^{mn}_{\bf p}|_{K} = g^{mn}_{\bf p}|_{K'}=g^{-m,-n}_{\bf p}$ for $u=0$,
and use the completeness relation~\cite{KSLs} 
\begin{equation}
\sum_{n=-\infty}^{\infty} |g^{mn}_{\bf p}|^{2} = e^{\ell^{2}{\bf p }^{2}/2}
\label{compRel}
\end{equation}
to extend the sum $\sum_{n\in {\rm DS}}$ to its complement 
$\sum_{n\in \overline{\rm DS}}$ as well.
The result is 
\begin{equation}
\sum_{n \in {\rm DS}} |g^{jn}_{\bf p}|^{2} 
\stackrel{u=0}{=} {1\over{2}}\, (e^{\ell^{2}{\bf p }^{2}/2} - |g^{j0}_{\bf p}|^{2}
- |g^{j1}_{\bf p}|^{2}- |g^{j2}_{\bf p}|^{2}),
\label{DSsum}
\end{equation}
for $j\in (0,1,2)$.
Equation~(\ref{compRel}) was noted earlier
with a formal proof;
a direct proof of it is given in Appendix A.
The $e^{ \ell^{2}{\bf p }^{2}/2}$ term in Eq.~(\ref{DSsum}),
though leading to a divergence upon integration over ${\bf p}$,
is common to all levels $j$ and is safely omitted.
We thus take the rest as
the regularized expression for $\sum_{n \in {\rm DS}} |g^{jn}_{\bf p}|^{2}$.

The regularized Dirac-sea contribution thus reads  
\begin{eqnarray}
&&V^{\rm DS}_{\rm X}
\stackrel{u\rightarrow 0}{=} 
\epsilon^{\rm v}_{0}\, R^{00;aa}_{\alpha\alpha;{\bf 0}}
+ \epsilon^{\rm v}_{1}\, R^{11;aa}_{\alpha\alpha;{\bf 0}}
+ \epsilon^{\rm v}_{2}\, R^{22;aa}_{\alpha\alpha;{\bf 0}}, \\
&&\epsilon^{\rm v}_{j} = {\textstyle{1\over{2}}} 
\sum_{\bf p}v_{\bf p}\gamma_{\bf p}^{2}\, 
 \sum_{n=0}^{2} |g^{jn}_{\bf p}|^{2}.
\label{VxDStwo}
\end{eqnarray}
Integration over ${\bf p}$, with the aid of  the formula 
\begin{equation}
\sum_{\bf p}v_{\bf p}\gamma_{\bf p}^2 \,[1, q^2, q^4, q^6, q^8] 
= [1,1,3,15,105]\, \tilde{V}_{c} 
\end{equation}
with $q\equiv \ell\, |{\bf p}|$, 
then yields
\begin{eqnarray}
\epsilon^{\rm v}_{0} &=& 
\textstyle{1\over{2}}\,\big[ 1 
+  c_{1}^2 c_{2}^2\, ({7\over{8}} + {11\over{8}}\, \kappa^2 
+ \kappa^4) \big]\, \tilde{V}_{c},
\nonumber\\
\epsilon^{\rm v}_{1} &=& 
\textstyle{1\over{2}}\,\big[ 1 
+  c_{1}^4 c_{2}^2\, ({11\over{16}} + {15\over{16}}\, \kappa^2 
+ \kappa^4) \big]\, \tilde{V}_{c},
\nonumber\\
\epsilon^{\rm v}_{2} &=& 
\textstyle{1\over{2}}\,\big[ 1 
+  c_{1}^2 c_{2}^4\, ({29\over{64}} - {7\over{64}}\, \kappa^2 
-{11\over{8}} \kappa^4 -{15\over{4}} \kappa^6 - 2\kappa^8  ) \big]\, 
\tilde{V}_{c},
\nonumber\\
\label{Vzeroreg}
\end{eqnarray}
where  $c_{1}^2\equiv (c_{1})^2$, etc.,  
$\tilde{V}_{c} \equiv \sqrt{\pi/2}\, V_{c}$
and 
\begin{equation}
V_{c} \equiv \alpha/(\epsilon_{b}\ell) \approx (56.1/\epsilon_{b})\, 
\sqrt{B[{\rm T}]}\, {\rm meV}.
\end{equation}
Numerically,
\begin{eqnarray}
(\epsilon^{\rm v}_{0}, \epsilon^{\rm v}_{1}, \epsilon^{\rm v}_{2}) 
= (0.888, 0.777, 0.641) \,  \tilde{V}_{c}
\label{Evn} 
\end{eqnarray}
for $\hat{\gamma} = 1/\kappa \approx 3.41$ at $B=10$ T.

Vacuum fluctuations thus shift the $n=0, 1$ and $2$ modes differently 
and the splitting among 
$(\epsilon^{\rm v}_{0}, \epsilon^{\rm v}_{1}, \epsilon^{\rm v}_{2})$
reflects the difference in their spatial distributions, 
as is clear from Eq.~(\ref{VxDStwo}). 
The empty PZM levels are ordered as 
$\epsilon^{\rm v}_{0} > \epsilon^{\rm v}_{1} >\epsilon^{\rm v}_{2}>0$.
Actually the spectra vary with filling of the PZM sector.
Note Eq.~(\ref{VxDS}), which tells us to include extra contributions
$-|g^{jn}|^2$  
for $\epsilon^{\rm v}_{j}$, when the $n\in (0,1,2)$ level is filled.
In particular,  when the PZM sector is filled up, one finds that   
$\{\epsilon^{\rm v}_{j}\}$ change sign so that they are ordered as
$\epsilon^{\rm v}_{0} < \epsilon^{\rm v}_{1} <\epsilon^{\rm v}_{2} < 0$.

Let us next suppose 
filling the lowest-lying $n=2$ level first in the empty PZM sector 
(for $u= \mu_{\rm Z}=0$).
One then finds $\epsilon^{\rm v}_{2} \approx 0.054\, \tilde{V}_{c}$
for $g \approx 3.41$.
If, instead, the highest-lying $n=0$ level were first filled, 
one would find
$\epsilon^{\rm v}_{0} \approx -0.11\, \tilde{V}_{c}$.
This puzzling situation suggests that one cannot reach 
the true ground state by filling the $n=2$ level alone.
It is clear now that one has to diagonalize 
the exchange interaction~(\ref{Vex}), 
with mixing among the $n=(0,1,2)$ orbital modes taken into account.

\section{mixing of the PZM levels}

In this section we examine how the PZM sector changes in spectrum with filling.
The first step is to extract from $V_{X}$ in Eq.~(\ref{Vex}) the exchange interaction 
acting within the $n=(0,1,2)$ sector,   
\begin{eqnarray}
V_{\rm X}^{\rm pz}&=&
-  \sum_{\bf p}v_{\bf p}\gamma_{\bf p}^{2}\,  \Gamma^{nm}_{\bf p}\, R^{mn}_{\bf 0},
\nonumber\\
\Gamma^{00}_{\bf p} &=& \nu^{nn} |g^{n0}_{\bf p}|^{2},
\Gamma^{11}_{\bf p} = \nu^{nn} |g^{n1}_{\bf p}|^{2},
\Gamma^{22}_{\bf p} = \nu^{nn} |g^{n2}_{\bf p}|^{2}, \nonumber\\
\Gamma^{10}_{\bf p} &=& g^{00}_{\bf p} g^{11}_{\bf -p}\,  \nu^{10} 
+ g^{01}_{\bf p}  g^{21}_{\bf -p}\,  \nu^{21},\nonumber\\
\Gamma^{21}_{\bf p} &=& g^{10}_{\bf p} g^{12}_{\bf -p}\,  \nu^{10} 
+ g^{11}_{\bf p}  g^{22}_{\bf -p}\,  \nu^{21},\nonumber\\
\Gamma^{20}_{\bf p} &=& g^{00}_{\bf p} g^{22}_{\bf -p}\,  \nu^{20},
\label{Vxpz}
\end{eqnarray}
where  $m, n\in (0,1,2)$; 
$\Gamma^{01}_{\bf p} \equiv (\Gamma^{10}_{\bf -p})^{*}$,
$\Gamma^{21}_{\bf p} \equiv (\Gamma^{12}_{\bf -p} )^{*}$, etc.
For conciseness the spin and valley indices have been suppressed in the above;
$\nu^{nn}\,  R^{00}_{\bf 0}$, e.g., reads
$\nu^{nn;ba}_{\beta \alpha}\,  R^{00;ab}_{\alpha \beta; {\bf 0}}$.

Let us, for the moment, freeze the spin and valley degrees of freedom 
and focus on the orbital degrees of freedom.
The PZM sector then consists of three levels $n\in (0,1,2)$ 
governed by the effective Hamiltonian 
${\cal V}\equiv V^{\rm DS}_{\rm X} + V_{\rm X}^{\rm pz}
= H^{mn} R^{mn}_{\bf 0}$
with 
$H^{mn} = \epsilon_{n}^{\rm v}\,  \delta^{mn} 
-  \sum_{\bf p}v_{\bf p}\gamma_{\bf p}^{2}\,  \Gamma^{nm}_{\bf p}$.
Note that $\Gamma^{mn}_{\bf p}$ are real for real filling factors $\nu^{mn}$, 
which we take.
It therefore suffices to use a real $O(3)$ rotation,
rather than a full $SU(3)$ rotation,
to diagonalize the $3 \times 3$ real symmetric matrix  $H^{mn}$.
We thus rotate 
$\psi^{m} =(\psi^{0}, \psi^{1}, \psi^{2})$ in orbital space,
\begin{eqnarray}
\psi^{m} (y_{0}) &=&  
[{\cal U}(\theta_{2},\theta_{1},\theta_{0})]^{m n}\, \Phi^{n} (y_{0}),
\label{rotatepsi}
\end{eqnarray}
with  three Euler angles $(\theta_{2},\theta_{1},\theta_{0})$ 
parameterizing
\begin{equation}
{\cal U}(\theta_{2},\theta_{1},\theta_{0})
= e^{i \theta_{2}\, t_{2}} e^{i \theta_{1}\, t_{1}}  e^{i \theta_{0}\, t_{0}},
\end{equation}
where the spin-1 generators $(t_{a})^{bc} \equiv i \epsilon^{bac}$ 
in terms of the totally antisymmetric tensor 
$\epsilon^{abc}$ with $ \epsilon^{012} =1$.
Note that $\theta_{0}$ mixes $n=(1,2)$, $\theta_{1}$ mixes $(0,2)$, etc.

Via the rotation, ${\cal V}= H^{mn} R^{mn}_{\bf 0}
= {\cal H}^{mn} {\cal R}^{mn}_{\bf 0}$
with ${\cal H} = {\cal U}^{\dag} H\,  {\cal U}$,
where
${\cal R}^{mn}_{\bf 0}$ stand for the charge operators 
for $\Phi^{n}$, i.e., 
$R^{mn;ab}_{\alpha\beta;{\bf 0}}$ with 
$\psi^{n;b}_{\beta}\rightarrow \Phi^{n;b}_{\beta}$.
The transformed fields $\Phi^{n}$ are taken to diagonalize ${\cal H}^{mn}$ 
and hence the associated filling factors as well, 
$N_{n} \propto \langle G|(\Phi^{n})^{\dag} \Phi^{n}|G\rangle$ 
with $0 \le N_{n} \le 1$ and $n \in (0,1,2)$;
one can now write  $\nu^{mn} =(U_{mn'} )^*  N_{n'}  (U^t)_{n'n} $.

Let us start filling the empty PZM sector at 
(relative) filling factor $n_{\rm f}=0$.
Obviously, in view of  level splitting~(\ref{Evn}),
it is the lowest-lying $n=2$ level ($\Phi^{2}$)
that starts to be filled.
To follow how it evolves let us suppose 
that it is filled with fraction  $n_{\rm f}\le 1$ 
and substitute $(N_{0}, N_{1}, N_{2}) = (0,0,n_{\rm f})$.
${\cal H}^{mn}$ is diagonalized if one can adjust 
$(\theta_{0},\theta_{1},\theta_{2})$ so that 
${\cal H}^{01} = {\cal H}^{02}= {\cal H}^{12}=0$.

Note first that, with no level mixing, 
i.e., $\theta_{0}=\theta_{1}=\theta_{2}=0$, 
the eigenvalues $\{{\cal H}^{nn}\}$ simply go down 
with increasing $n_{\rm f}$.
Note next that,  to first order in $\{ \theta_{n} \}$,  
\begin{eqnarray}
{\cal H}^{12} &\approx&  ( 0.136 - 0.268\,  n_{\rm f})\,  \theta_{0} + ...,\nonumber\\ 
{\cal H}^{02} &\approx& - (0.247 - 0.130\, n_{\rm f})\,  \theta_{1}  + ..., \nonumber\\ 
{\cal H}^{01} &\approx& -0.201\, n_{\rm f}\,  \theta_{0}  
+ ( 0.111 + 0.0626\,  n_{\rm f})\,   \theta_{2}  + ... \ \ .
\end{eqnarray}
This structure reveals that 
$\theta_{0}=\theta_{1}=\theta_{2}=0$ for $n_{\rm f}< n_{\rm cr}\approx 0.507$ 
while $\theta_{0}\not=0$ is possible for $n_{\rm f} > n_{\rm cr}$.
Solving for   $\{ \theta_{n} \}$ numerically for $n_{\rm f} \ge n_{\rm cr}$
shows that the energy eigenvalue ${\cal H}^{22}$ 
is indeed lowered for $n_{\rm f} > n_{\rm cr}$
with $ \theta_{0} \not=0$.
One can then reach the $n_{\rm f}=1$ state, 
and setting 
$(N_{0}, N_{1}, N_{2}) \rightarrow (0,n _{\rm f}-1,1)$,
etc., takes one further to the $n_{\rm f}=$2 and 3 states.


\begin{figure}[tbp]
\includegraphics[scale=0.85]{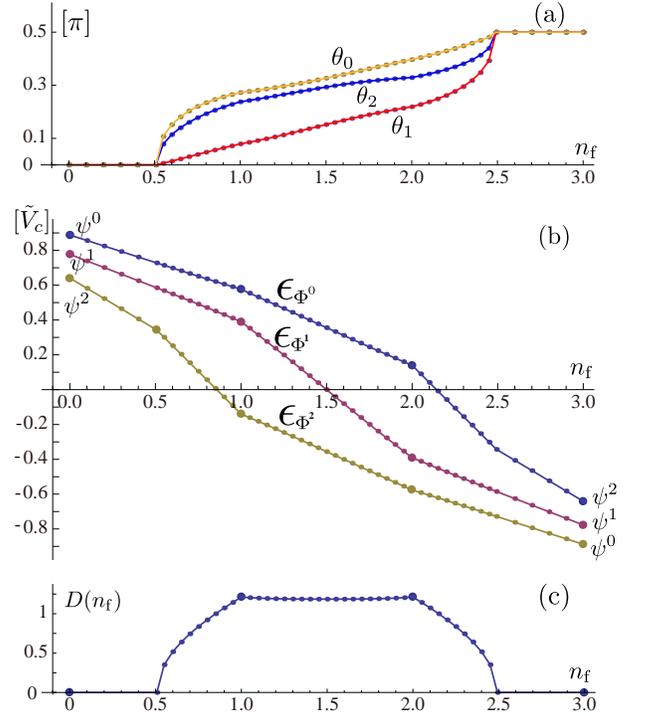}  
\caption{
Orbital mixing.
(a) Angles $(\theta_{0}, \theta_{1},\theta_{2})$ vary from 0 to $\pi/2$
with filling of the $(\Phi^{0}, \Phi^{1}, \Phi^{2})$ sector.
(b) Variations of the spectra of the $\Phi^{0}, \Phi^{1}$ and $\Phi^{2}$ levels
over the range of relative filling factor $n_{\rm f} \in [0,3]$.
(c) Electric dipole moment induced via orbital mixing.
}
\end{figure}


Figure 1 (a) shows how angles $\{ \theta_{n} \}$ vary as 
$n_{\rm f}$ is increased from 0 to 3.
Actually we find another solution 
which differs from one shown in the figure by signs, 
$(\theta_{2}, \theta_{1}, \theta_{0}) 
\rightarrow (-\theta_{2}, \theta_{1}, -\theta_{0})$.
These two solutions are related by a unitary transformation
$Y= {\rm diag}[-1,1,-1]$, 
with ${\cal U}(-\theta_{2}, \theta_{1}, -\theta_{0})  
=Y {\cal U}(\theta_{2}, \theta_{1}, \theta_{0}) Y^{-1}$,
so that $Y\psi = {\cal U}(-\theta_{2}, \theta_{1}, -\theta_{0})Y\Phi$.
They naturally lead to the same level spectra $\{\epsilon_{\Phi^{n}} \}$ 
depicted in Fig.~1 (b).

In Fig.~1 (a), each $\theta_{n}$ evolves 
from 0 to $\pi/2$ with increasing $n_{\rm f}$. 
The eigenmodes
$(\Phi^{0}, \Phi^{1}, \Phi^{2})$ thereby continuously change from
$(\psi^{0}, \psi^{1}, \psi^{2})$ to  $(\psi^{2}, -\psi^{1}, \psi^{0})$.
The  empty $n=(0, 1,2)$ levels at $n_{\rm f}=0$ 
thus turn into the filled $n=(2, 1, 0)$ levels of energies 
$(-|\epsilon^{\rm v}_{2}|,-|\epsilon^{\rm v}_{1}|,-|\epsilon^{\rm v}_{0}|)$,
respectively, at  $n_{\rm f}=3$, without any level crossing. 
Each spectrum $\epsilon_{\Phi^{n}}(n_{\rm f})$ 
goes down with $n_{\rm f}$, 
with marked change across $n_{\rm f} \sim (0.51, 1, 2, 2.49)$.  
The spectra as a whole  realize particle-hole symmetry, with
$\epsilon_{\Phi^{0}}(n_{\rm f}) =- \epsilon_{\Phi^{2}}(3- n_{\rm f})$ and 
$\epsilon_{\Phi^{1}}(n_{\rm f}) =- \epsilon_{\Phi^{1}}(3- n_{\rm f})$
in obvious notation.
In particular, the band gaps at $n_{\rm f}=$1 and 2 are equal, with  
\begin{equation}
\epsilon^{\rm gap}|_{n_{\rm f}=1,2} 
\approx  0.53\, \tilde{V}_{c},
\label{orbitalGap}
\end{equation}
considerably smaller than the full Coulombic gap 
$2\, \epsilon_{2}^{v} \approx  1.28\, \tilde{V}_{c}$.

A special feature associated with orbital mixing is 
that charge carriers acquire electric dipole
moment, as noted earlier~\cite{KSpzm, CLBM} for bilayer graphene.
To see this let us consider coupling to an external scalar potential $A_{0}$,
with the Hamiltonian
$H_{A} = - e\sum_{\bf p} (A_{0})_{\bf p}\, \rho_{\bf -p}$.
Note that 
$g^{01}_{\bf p}, g^{12}_{\bf p}\, ({\rm in}\ \rho_{\bf -p}) \propto p$, 
which implies that orbital mixing gives rise to coupling 
to an inplane electric field
${\bf E}_{\parallel}=(E_{x}, E_{y}) = - \nabla A_{0}$.
Indeed, for a spatially almost uniform field ${\bf E}_{\parallel}$, 
the relevant portion of $H_{A}$ 
is written as  $H_{A} \approx h_{A}^{mn}R_{\bf 0}^{mn}$
with 
 $h_{A} = -c_{2}\, (e\ell/\sqrt{2})\, (E_{y} T_{y} + E_{x}T_{x})$,
 where
\begin{equation}
T_{y} = \left(
\matrix{ & 1&   \cr 
1& & \lambda \cr
&  \lambda &  \cr}
\right), \ 
T_{x} =i \left(
\matrix{ & -1&   \cr 
1& & -\lambda \cr
&  \lambda &  \cr}
\right)
\end{equation}
act on fields $(\psi^{0}, \psi^{1}, \psi^{2})^{\rm t}$ and
$\lambda \equiv \lambda_{{\bf p}=0} = \sqrt{2} \, ( 1 + \kappa^2)\, c_{2}$.

The expectation value $\langle G| H_{A} |G \rangle$ then reads
\begin{equation}
\langle G| H_{A} |G \rangle \approx \rho_{0} \int d^{2}{\bf x}\, 
(- {\bf d}\cdot {\bf E}_{\parallel}),
\end{equation}
where 
${\bf d} = (0, d_{y})$ and $d_{y}= (e \ell/\sqrt{2})\, c_{1} D(n_{\rm f})$
with function 
$D(n_{\rm f}) \equiv (T_{y} + \lambda\,  T_{x})^{mn}\, \nu^{mn}$ 
given by the plot in Fig 1 (c).
This shows that electrons acquire electric dipole moment of magnitude 
$|{\bf d}_{e}|=  (e \ell/\sqrt{2})\, c_{1} |D(n_{\rm pz})|$
(per particle), pointing in the $y$ axis for the present choice of wave functions.

Actually the electric dipole can point in any direction 
(in general, perpendicular  to traveling waves)
at no cost of energy.
To see this let us consider a phase rotation of the form 
(within $SU(3)$ rotations),
$\psi^{n}\rightarrow \hat{\psi}^{n}$
with $(\psi^{0}, \psi^{1}, \psi^{2})
= (e^{-i\phi} \hat{\psi}^{0}, \hat{\psi}^{1},e^{i\phi} \hat{\psi}^{2})$.
Note that $g^{k n}_{\bf p}$ are thereby transformed 
so that $\rho_{-{\bf p}}$ remains invariant.
This transformation leaves
${\cal V} =V^{\rm DS}_{\rm X} + V_{\rm X}^{\rm pz}$ invariant; 
hence,
the spectrum remains unchanged.
Still the electric dipole thereby rotates so that  
\begin{eqnarray}
{\bf d}\equiv (d_{x}, d_{y})= (\sin \phi, \cos \phi)\,  |{\bf d}_{e}| .  
\end{eqnarray}
It is now clear that a pair of solutions $\psi$ and $Y\psi$, 
encountered earlier, differ by a rotation by $\pi$ of coordinates in the sample plane.

\section{Generalization}

In this section we recover the electron spin and valley degrees of freedom 
and explore the PZM sector with 
both $\mu_{\rm Z}$ and bias $u$,
using  the full  Hamiltonian 
\begin{equation}
H_{\rm eff}= H_{u} + V_{X}^{\rm DS} + V_{X}^{\rm pz} .
\end{equation}
We leave $u$ arbitrary but keep $|u| \ll V_{c}$
so that one can still use the $u=0$ expressions 
for ${\cal V}= V_{X}^{\rm DS} + V_{X}^{\rm pz}$,
with $u$ retained  only in $H_{u}$ 
as a small perturbation.~\cite{fnlargeu}

In addition, we ignore the difference between 
the intra- and interlayer Coulomb potentials 
that leads to a valley-symmetry breaking 
of  $O(V_{c}\, d/\ell)$, with the layer separation $d \sim 0.35$ nm $\ll \ell$. 
This breaking contains capacitance energies that determine how valleys rotate.
In conventional bilayer systems, the capacitance energy, though
as tiny as $O(V_{c}\, d^2/\ell^2)$, is positive and induces a valley rotation 
$(K,K') \rightarrow K \pm K'$, which makes 
the symmetric states $\propto K+K'$ lower in energy. 
In contrast, for bilayer graphene, capacitance energies 
turn out to be negative~\cite{KSLs} 
and suppress possible valley rotations for $u \sim 0$.

Experimentally, it is difficult to directly observe valley quantum numbers,
especially from the sequence in which the broken-symmetry states 
emerge with varying filling factor $\nu$ or magnetic field $B$.
The sequence is governed by the Coulombic gaps, 
which, though possibly triggered by small valley or spin or orbital breaking,
are practically insensitive in magnitude to small $|u| \ll V_{c}$.
(In contrast, for large bias $u$, the valley is naturally polarized 
in either $K$ or $K'$, depending on the sign of $u$.)
For this reason, 
instead of a (rather laborious) analysis of capacitance energies,
we here simply suppose a possible valley rotation
$(K,K') \rightarrow (+,-)$ without specifying its details for small $u$;
we take the (-)  state to be lower in energy 
for each $n\in (0,1,2)$ and spin $\alpha \in (\uparrow, \downarrow)$.


\begin{figure}[tbp]
\includegraphics[scale=0.95]{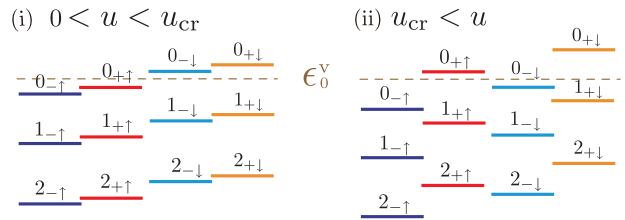}  
\caption{
Empty levels in the PZM sector at $\nu=-6$;
for illustration, (i) $u = 0.25\, u_{\rm cr}$ and (ii) $u= 1.4\, u_{\rm cr}$ 
with $u_{\rm cr} \equiv \mu_{\rm Z}/(1-z_{2}) \approx 1.2\, \mu_{\rm Z}$
and $\mu_{\rm Z}/\tilde{V}_{c}=0.05$.
}
\end{figure}


In $H_{\rm eff}$ the exchange interaction 
$V^{\rm DS}_{\rm X} + V_{\rm X}^{\rm pz}$ 
conserves both valley and spin, but breaks the orbital degeneracy.
In contrast, the small perturbation $H_{u}$
lifts all three degeneracies.
Figure 2 depicts the empty PZM 
sector (at $\nu=-6$) 
governed by $H_{u} + V^{\rm DS}_{\rm X}$,
with level spectra
\begin{equation}
\epsilon^{ \pm \uparrow}_{n}
= \epsilon_{n}^{\rm v}  \pm \epsilon_{n}^{u} 
- \textstyle{1\over{2}}\mu_{\rm Z}, \ \
\epsilon^{ \pm \downarrow}_{n}
=\epsilon_{n}^{\rm v} \pm \epsilon_{n}^{u} 
+ \textstyle{1\over{2}}\mu_{\rm Z} ,
\label{emptyPZM}
\end{equation}
in obvious notation. 
There are two possible level patterns, 
depending on (i) $0\le u < u_{\rm cr}$
(of spin-breaking domination) 
or (ii) $u > u_{\rm cr}$ (of valley-breaking  domination)
with 
$u_{\rm cr} \equiv \mu_{\rm Z}/(1-z_{2}) \, 
[\approx 1.2\, \mu_{\rm Z}$ at 10T]. 
In Eq.~(\ref{emptyPZM}), for definiteness, 
we have assumed no valley rotation and $u\ge 0$,
so that $(+,-) = (K,K')$.
When a valley rotation is induced (for small $u$), 
the $\pm \epsilon_{n}^{u}$ portions are replaced by more complex expressions,
which, at any rate, are small for small $u$, and the level pattern (i) in Fig.~2
remains essentially intact.  (For consistency, we set $u \rightarrow 0$ 
in our discussion for case (i) below.)


\begin{figure*}[btp]
\includegraphics[scale=0.8]{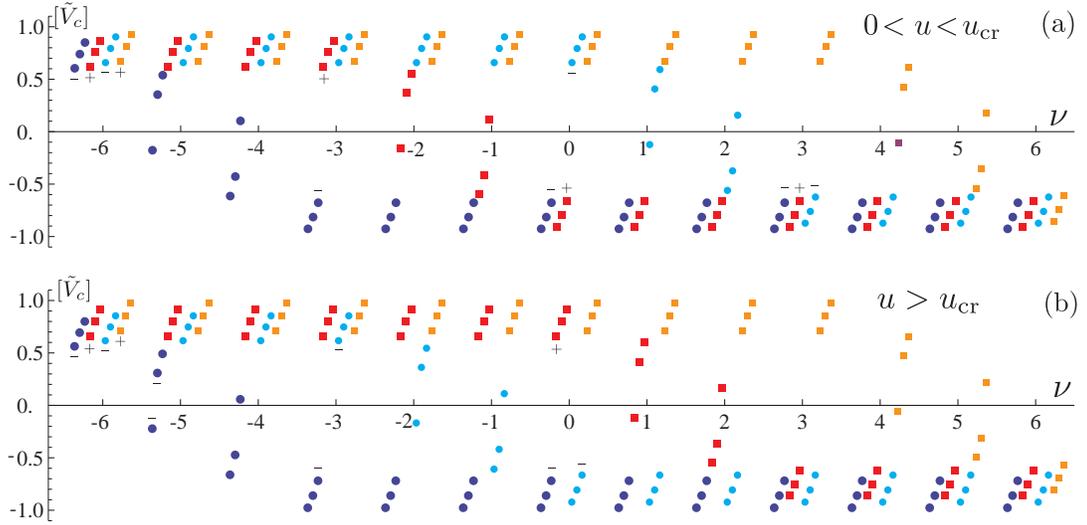}
\caption{
Spectra of the PZM Landau levels 
at each integer filling factor $\nu \in [-6,6]$.
(a) $u=  0.4\, \mu_{\rm Z}  <  u_{\rm cr}$ 
and (b) $u=2\, \mu_{\rm Z} > u_{\rm cr}$ 
with $\mu_{\rm Z}/\tilde{V}_{c} =0.05$
and $u_{\rm cr} \equiv \mu_{\rm Z}/(1-z_{2}) \approx 1.2\, \mu_{\rm Z}$
for illustration.
Large blobs and squares refer to spin-up levels 
$(2_{-\uparrow},1_{-\uparrow}, 0_{-\uparrow})|_{\theta}$ 
and $(2_{+\uparrow},1_{+\uparrow},0_{+\uparrow})|_{\theta}$, 
respectively, from left to right; smaller symbols refer to
those for the spin-down levels. 
Here $(2_{-\uparrow},1_{-\uparrow}, 0_{-\uparrow})|_{\theta}$, 
e.g., stands for 
$(2_{-\uparrow},1_{-\uparrow}, 0_{-\uparrow})$  
for empty levels $(\theta=0)$ and 
$(0_{-\uparrow},1_{-\uparrow}, 2_{-\uparrow})$ 
for filled levels $(\theta=\pi/2)$,
in accordance with Fig.~1;
empty levels have positive energy and occupied levels 
have negative energy.
Valley indices $\pm$ are attached to some symbols 
to indicate the nature of the associated gaps.
}
\end{figure*}


Let us start filling the empty PZM sector.
Obviously,  with $|u|, \mu_{\rm Z} \ll V_{c}$, the orbital splitting
among $\{ \epsilon_{n}^{\rm v} \}$ 
singles out the $2_{-\uparrow}$ level as the lowest-lying one    
in both cases (i) and (ii). 
It is thus the $2_{-\uparrow}$  level that is filled first. 
As it is being filled, 
it comes down in energy, followed by the $0_{-\uparrow}$ and
$1_{-\uparrow}$ levels
coupled via the exchange interaction $V_{X}^{\rm pz}$. 
These three levels undergo orbital mixing, discussed in the previous section, 
through the $\nu=-5$ and -4 states 
until one reaches the $\nu=-3$ state, 
which is orbitally neutral (an $SU(3)$ singlet) but
is polarized in valley and spin $(-, \uparrow)$.
The associated $\nu=-3$ level gap is
a valley gap for case (i) and a spin gap for case (ii), 
\begin{eqnarray}
\epsilon^{\rm gap}_{\nu= -3}|_{\rm (i)} 
&\approx&  
2\, \epsilon_{2}^{\rm v} ,
\nonumber\\
\epsilon^{\rm gap}_{\nu= -3}|_{\rm (ii)}
&=&  2\, \epsilon_{2}^{\rm v} + \mu_{\rm Z},
\label{Egapnuthree}
\end{eqnarray}
with $ 2\, \epsilon_{2}^{\rm v} \approx 1.28\, \tilde{V}_{c}$.
Similarly, as one goes up from $\nu=-3$ to $\nu=0$, 
essentially the same orbital mixing is repeated for the $(+,\uparrow)$ sector 
in case (i) and for the $(-,\downarrow)$ sector in case (ii);
analogously for the $\nu \in [0,6]$ domain.


Figures 3 (a) and 3 (b) show the resulting spectra of the PZM multiplet
at each integer filling factor $\nu \in [-6, 6]$.
They differ in pattern 
for (i) $u\sim 0$ and (ii) $u > u_{\rm cr}$,
but form a perfectly particle-hole symmetric spectrum
for the PZM sector  in each case.
The $\nu=\pm 2, \pm 1$ and 0 states thus differ in composition, 
depending on $u$. 
The $\nu=0$ state, in particular, is spin-polarized for $u \sim 0$
and valley-polarized for $u > u_{\rm cr}$, with a gap
\begin{eqnarray}
\epsilon^{\rm gap}_{\nu= 0}|_{\rm (i)} &\approx&  
2\, \epsilon_{2}^{\rm v} + \mu_{\rm Z} ,
\nonumber\\
\epsilon^{\rm gap}_{\nu= 0}|_{\rm (ii)} 
&=&  2\, \epsilon_{2}^{\rm v} + (1 - z_{2})\, u -\mu_{\rm Z}.
\label{Egapnuzero}
\end{eqnarray}

As to the $\nu=-5$ gap, especially for case (ii) (of relatively large $u$), 
we note the following:
$(\epsilon_{0}^{u}, \epsilon_{1}^{u}, \epsilon_{2}^{u})$ in $H_{u}$,
via the rotation ${\cal U}$, 
turns into $(0.96, 0.899, 0.897)\, u/2$, 
i.e., $\epsilon_{1}^{u} \approx \epsilon_{2}^{u}$ at $\nu=-5$;
similarly,  $\epsilon_{0}^{u} \approx \epsilon_{1}^{u}$ at $\nu=-4$.
This suggests that the $\nu=(\pm 5, \pm 4, \pm 2, \pm 1)$ gaps are 
practically insensitive to both bias $u$ and $\mu_{\rm Z}$, and 
equal to $\epsilon^{\rm gap}|_{n_{\rm f}=1,2} $ 
in Eq.~(\ref{orbitalGap}),
\begin{equation}
\epsilon^{\rm gap}_{\nu= \pm 5, \pm 4, \pm 2, \pm 1}
\approx 0.53\, \tilde{V}_{c} .
\label{Egapnufive}
\end{equation}
These orbital gaps are considerably smaller 
than the (Coulomb-enhanced $\nu=0, \pm 3$) spin or valley gaps, 
\begin{equation}
\epsilon^{\rm gap}_{\nu= \pm 1,\pm 2,\pm 4, \pm 5} 
< \epsilon^{\rm gap}_{\nu= \pm 3} 
\lesssim \epsilon^{\rm gap}_{\nu=0}\ ( \ll \epsilon^{\rm gap}_{\nu= \pm 6}),
\end{equation}
in conformity with Hund's rule.~\cite{BCNM, ZTM}
These $\nu = \pm 1,\pm 2, ...$ orbital gaps 
and the $\nu=\pm 3$ valley gaps 
for $u\sim 0$ barely depend on $\mu_{\rm Z}$ and 
will therefore be insensitive to an additional parallel field $B_{\parallel}$
in experiments with a tilted magnetic field,
in contrast to the $\nu=0$ spin gap for $u\sim 0$.
For bilayer graphene, the corresponding gaps
take place at $\nu=-3$ and -2, 
and it was observed~\cite{ZCZJ} 
that the associated resistance minima are barely affected 
by $B_{\parallel}$ .

The orbitally polarized states at $\nu= \pm 1,\pm 2,\pm 4, \pm 5$ 
have spontaneous electric dipole moment 
and may potentially be unstable~\cite{KSpzm, BCR,CRGB} 
against charge inhomogeneities.
Their spectra may be modified 
(in random patterns or regular~\cite{BCR,CRGB} patterns)
around local charge concentrations but, 
as long as the orbital gaps survive, the quantum Hall states would emerge.
Such an instability disappears when bias $u$ is 
sufficiently large to stabilize 
the valley-polarized states.
For bilayer graphene full splitting of the PZM levels 
has indeed been observed.~\cite{FMY,ZCZJ}

The transport properties  of trilayers have been studied 
in a number of 
experiments.~\cite{BZZL,KEPF,TWTJ,LLMC,BJVL,ZZCK,JCST}
Experimentally there is clear evidence 
for formation of the quantum Hall states 
in the basic filling-factor sequence 
$\nu= \pm 4(N +3/2) = \pm6, \pm 10,\pm 14,...$ 
for both $ABC$-and $ABA$-stacked trilayers.
Evidence is yet very limited for the fine structure of the PZM sector
 with $|\nu| < 6$ in $ABC$ trilayers:
 An experiment,~\cite{ZZCK}
 using a Hall-bar device, observed a weak anomaly in $\sigma_{xy}$ 
indicative of the developing $\nu = \pm 3$ gap. 
A clear signal for the $\nu=0$ gap comes from 
the observation~\cite{BZZL,BJVL} 
of the insulating state at the Dirac point  ($\nu=0$) in $ABC$-trilayer devices, 
both suspended and substrate-supported ones, 
with the resistance rising exponentially 
with increasing $B$ and lowering temperature $T$.    
Experimentally, it is normally the $\nu=0$ insulating state 
that is first observed as a nontrivial feature 
within the PZM sector of few-layer graphene.
This suggests that the $\nu=0$ gap is an interaction-enhanced gap
rather than the far smaller intrinsic spin or valley gap.
The $\nu=\pm3$ gaps will be the next to be visible via quantized conductance.
In view of Eqs.~(\ref{Egapnuthree}) and (\ref{Egapnuzero}),
the $\nu=0$ gap will become even more prominent with increasing bias $u$,
in contrast to the $\nu =\pm 3$ gap.

Finally we wish to discuss possible effects of 
nonleading interlayer couplings
$(v_{4}, v_{3}, \gamma_{2})$.
The effect of $v_{4}$ can be included 
in ${\cal H}_{n}$ of Eq.~(\ref{Hn}) 
while $v_{3}$ and $\gamma_{2}$ induce transitions 
that go outside the PZM sector, 
as seen from the solutions in Eqs.~(\ref{psizero}) - (\ref{psitwo}).
Accordingly the spectra $(\epsilon_{0},\epsilon_{1}, \epsilon_{2})$
 are corrected to first order in $v_{4}/v$ and to second order 
 in $v_{3}/v$ and $\gamma_{2}/\gamma_{0}$.  
With typical values~\cite{TWTJ,Malard}  
$v_{4}/v\equiv r_{4} \sim 0.01$,
$v_{3}/v =\gamma_{3}/\gamma_{0} \sim 0.1$, 
and $\gamma_{2} \sim -0.02$ eV, 
such corrections are generally small.
The leading $O(v_{4})$ corrections, in particular, 
may conveniently be included in $H_{u}$
if one sets
$\epsilon^{u}_{0} =u/2$, 
$\epsilon^{u}_{1} =(1-z_{1}) ( u/2 +2\kappa r_{4}\, \omega_{c})$ and  
$\epsilon^{u}_{2} =(1-z_{2})  u/2 +4\kappa (1 + \kappa^2)\, r_{4}\, \omega_{c}$,
with $2\kappa r_{4} \sim 0.005$ and $4\kappa (1 + \kappa^2)\, r_{4} \sim 0.01$. 
Unlike $u$, such $O(v_{4})$ corrections are common to 
the $K$ and $K'$ valleys and lead to weak electron-hole asymmetry.
The relative magnitude of 
$(\epsilon^{u}_{0},\epsilon^{u}_{1}, \epsilon^{u}_{2})$
may vary with bias $u$ and can potentially control a valley rotation 
for small $u \sim 0$. 
Still the orbital splitting among $\{\epsilon^{\rm v}_{n}\}$ 
is generally larger than the splitting among $\{\epsilon^{u}_{n}\}$, 
and the PZM sector will essentially maintain the spectra shown in Fig.~3.
The electron-hole symmetric spectra there will also serve as the base point 
for further examining possible effects of 
nonleading intra- and interlayer parameters.

\section{Summary and discussion}

In a magnetic field graphene trilayers acquire, on topological grounds,  
a special multiplet of nearly-zero-energy Landau levels
with a threefold degeneracy in Landau orbitals.
In this paper we have studied the structure of this PZM multiplet
in $ABC$-stacked trilayer graphene
and pointed out that its orbital degeneracy
is lifted by quantum fluctuations of the valence band.
Here we encounter a trilayer generalization of the $\lq\lq$orbital" Lamb shift,
discussed earlier for bilayer graphene.
The splitting among the shifted energies $\{\epsilon^{\rm v}_{n}\}$ 
acts as a quantum orbital breaking 
that generally exceeds intrinsic spin or valley breaking
in scale, and essentially governs the structure of the PZM sector.

The orbital Lamb shift of the PZM Landau levels 
is a $\lq\lq$field-theoretic" vacuum effect 
but is intimately correlated with the Coulomb interaction 
acting within the multiplet.
This is because they have to combine to yield 
an electron-hole symmetric spectrum for the PZM multiplet
(with only the leading couplings $\gamma_{0}$
and $\gamma_{1}$ kept)  as a whole.
In particular, large Coulombic gaps, expected at $\nu=0$ and $\pm 3$,  
are essentially given by the energy scale 
$\sim 2 \epsilon_{2}^{\rm v}$ 
of the orbital Lamb shift.

The PZM levels get mixed via the Coulomb interaction and avoid 
level crossing, 
keeping smaller orbital gaps (of magnitude $\sim  0.5\, \tilde{V}_{c}$),
as we have seen in Sec.~V.
Level crossing, if present, would enhance the degree of degeneracy 
and the steps of Hall plateaus would jump accordingly. 
Observations of possible $\nu = \pm1,\pm 2, \pm4, \pm 5$ 
quantum Hall states in high-quality samples, 
such as suspended or BN-supported ones, under high magnetic fields, 
if achieved, would be direct evidence 
for the presence of orbital mixing without level crossing.
It is also possible, in principle, to detect the orbital gaps 
via cyclotron resonance 
within the PZM sector.~\cite{BCNM,KSLs}

In this paper we have focused on $ABC$-stacked trilayer graphene. 
We remark that our analysis and conclusion cannot simply 
be carried over to the case of $ABA$ trilayers, 
which lacks a direct link between the $K$- and $K'$-valley expressions 
[such as Eq.~(\ref{SHS})]
and which thus requires a separate analysis.~\cite{fnABA}

\acknowledgments

This work was supported in part by a Grant-in-Aid for Scientific Research
from the Ministry of Education, Science, Sports and Culture of Japan 
(Grant No. 24540270).

\appendix

\section{Derivation of Eq.~(\ref{compRel})}

In this appendix we present a proof of  
the completeness relation
$\sum_{n} |g^{mn}_{\bf p}|^{2} = e^{\ell^{2}{\bf p }^{2}/2}$
in Eq.~(\ref{compRel}).
A simpler version of it is the following:
\begin{equation}
\sum_{n=0}^{\infty} f_{\bf p}^{k n} f_{\bf -p}^{n m}
= e^{ \ell^2 {\bf p}^2/2}\, \delta^{k m}
\label{sumff}
\end{equation}
for integers $k,m \ge 0$, which is verified by use of the explicit form of  
$f_{\bf p}^{k n}$ in Eq.~(\ref{fknp}). We show that  Eq.~(\ref{compRel})
is essentially reduced to Eq.~(\ref{sumff}).

Let us first look at Eq.~(\ref{gkn})
and put the (orthonormal set of) six eigenvectors of ${\cal H}_{n}$ 
for each $n \in (3,4,...)$ into the orthogonal matrix 
$T_{n}= ({\bf v}_{n_{1}}, {\bf v}_{n_{2}},..., {\bf v}_{n_{6}})$
with $|n_{\sigma}| =n$ and $\sigma \in (1,2,...,6)$.
The first row of $T_{n}$ is $(b^{(1)}_{n_{1}}, b^{(1)}_{n_{2}}, ..., b^{(1)}_{n_{6}})
\equiv \{ b^{(1)}_{n_{\sigma}}\}$, the second row is $\{d^{(1)}_{n_{\sigma}}\}$, etc.
These row vectors also form an orthonormal basis.
This feature is also true for $n \in (0,1,2)$, 
except that $T_{n}$ has a smaller rank.

In taking the product $\sum_{n}g^{kn}_{\bf p}\, g^{nm}_{\bf -p}$ 
one may first sum over $n_{\sigma}$
for each fixed $n = |n_{\sigma}|$.
One thereby encounters inner products of  the row vectors such as 
$ \sum_{\sigma}\{ b^{(i)}_{n_{\sigma}}\}  \{ b^{(j)}_{n_{\sigma}}\} = \delta^{ij}$
and $ \sum_{\sigma}\{ b^{(i)}_{n_{\sigma}}\}  \{ d^{(j)}_{n_{\sigma}}\} =0$.
The remaining sum over $|n|$ is essentially reduced to 
formula~(\ref{sumff}) and one eventually finds that
$\sum_{n}g^{kn}_{\bf p}g^{nm}_{\bf -p} =e^{\ell^2 {\bf p}^2/2} \delta^{|k|, |m|}\,
 ({\bf v}_{k} \cdot  {\bf v}_{m})$,
which leads to Eq.~(\ref{compRel}).



\begin{thebibliography}{99}
\bibitem{fngz}  For $n \in (0,1,2)$,
$b^{(i)}_{n}$ are odd in $u$ while $d^{(i)}_{n}$ are even in $u$;
also $z_{1}$ and $z_{2}$ are functions of  $(\hat{u},\hat{\gamma}^2)$.
\bibitem{fnABA} A preliminary look into $ABA$ trilayers reveals 
that the orbital Lamb shift significantly differs between the $K$ and $K'$ valleys.
\end{thebibliography}
\end{document}